\documentclass[12pt]{iopart}
\usepackage{graphics}

\begin{document}
\bibliographystyle{unsrt}

\title[Magneto-transmission as a probe of Dirac fermions in bulk graphite]{Magneto-transmission as a probe of Dirac fermions in bulk graphite}

\author{M. Orlita$^{1,2,3}$, C. Faugeras$^{1}$, G. Martinez$^{1}$, D. K. Maude$^{1}$, M. L. Sadowski$^{1}$, J. M. Schneider$^{1}$, and M. Potemski$^{1}$}

\address{$^{1}$Grenoble High Magnetic Field Laboratory, CNRS, Grenoble, France\\
$^{2}$Institute of Physics, Charles University, Prague, Czech
Republic\\
$^{3}$Institute of Physics, v.v.i., ASCR, Prague, Czech Republic}
\ead{orlita@karlov.mff.cuni.cz}
\begin{abstract}
Far infrared magneto-transmission spectroscopy has been used to
probe ``relativistic'' fermions in highly oriented pyrolytic and natural
graphite. Nearly identical transmission spectra of those two materials are obtained,
giving the signature of Dirac fermions via absorption lines with an energy that
scales as $\sqrt{B}$. The Fermi velocity is evaluated to be
$\tilde{c}=(1.02\pm0.02)\times 10^6$~m.s$^{-1}$ and the pseudogap  at the $H$ point
is estimated to $|\Delta|<10$~meV.

\end{abstract}


\section{Introduction}

The gold rush, which started after the discovery of Dirac-like
particles in
graphene~\cite{NovoselovNature05,ZhangNature05,BergerScience06},
naturally resulted in an intensified interest in the properties of
bulk graphite with the apparent aim to prove directly the presence,
and to investigate the nature, of massless holes in the vicinity of
the $H$ point, where according to the standard band structure of
graphite, formulated by Slonczewski, Weiss and McClure (SMW),  these
relativistic fermions are
located~\cite{SlonczewskiPR58,McClurePR57}.

The first sign of such particles was reported by Toy \emph{et
al.}~\cite{ToyPRB77} via the $\sqrt{B}$-dependent features in their
magneto-reflection experiment, which is a direct fingerprint
of Dirac particles. Further evidence did not come earlier than in
the ``graphene age'' with the reinterpretation of the
Shubnikov-de Haas (SdH) and de Haas-van Alphen experiments. The analysis
of the phase of these quantum oscillations, carried out by
Luk'yanchuk and Kopelevich~\cite{LukyanchukPRL04,LukyanchukPRL06}
suggested the presence of normal (massive) electrons with a Berry
phase 0 and Dirac holes with a Berry phase $\pi$ in bulk graphite.
Another proof of particles with a linear dispersion around the H
point was offered by angular resolved photoemission spectroscopy
(ARPES) performed by Zhou \emph{et al.}~\cite{ZhouNatPhys06} and
later on also by Gr\"{u}neis \emph{et al.}~\cite{GruneisPRL08}. Both
massive and massless particles were identified in scanning
tunnelling spectroscopy (STS) by Li and Andrei~\cite{LiNatPhys07}
and recently also in far infrared (FIR)
magneto-transmission~\cite{OrlitaPRL08}.

Hence, many different experimental techniques indicate the presence
of Dirac holes in bulk graphite. Nevertheless,  the mutual
consistency of individual reports represents a serious difficulty,
which merits closer scrutiny. The analysis of quantum
oscillations~\cite{LukyanchukPRL04,LukyanchukPRL06} assumes that
``bulk'' graphite is a system composed of graphene single- and
few-layers~\cite{LukyanchukPC} and thus no crystal ordering along
the $c$-axis exists, which directly contradicts the commonly
accepted SWM model. On the other hand, the available ARPES results
are in a relatively good agreement with the SWM model, but are
characterized by a relatively low accuracy and probe the sample
surface only. The sensitivity to the sample surface only is also a
problem for STS experiments, which additionally failed to reveal
Dirac holes in earlier equivalent measurements~\cite{MatsuiPRL05}.

The available FIR experiments are also not trouble-free when
compared, as the Dirac holes were observed
in Refs.~\cite{ToyPRB77,OrlitaPRL08}, but not seen in recent
magneto-reflection experiment~\cite{LiPRB06}. In principle, these
optical experiments support the validity of SWM model, nevertheless
the relative strength of spectral features related to the $H$ point
in comparison to the $K$ point is not in agreement with theoretical
calculations of optical conductivity in magnetic
fields~\cite{KoshinoPRB08}. Very recently, the SWM model was found
by Kuzmenko \emph{et al.}~\cite{KuzmenkoPRL08} to be fully
sufficient to explain the temperature dependence of the reflectivity
of graphite. Note that the graphite band structure parameters
estimated using optical methods~\cite{ToyPRB77,OrlitaPRL08}, for
example the pseudogap at the $H$ point, differ from those obtained
from  ARPES experiments~\cite{GruneisPRL08}. A similar problem
occurs for the position of the Fermi level, as the hole density
inferred from the analysis of quantum
oscillations~\cite{LukyanchukPRL04,LukyanchukPRL06} is almost an order
of magnitude higher than that obtained from FIR
spectroscopy~\cite{OrlitaPRL08}.

\begin{figure}
\begin{center}
\scalebox{1.5}{\includegraphics*[13pt,9pt][240pt,174pt]{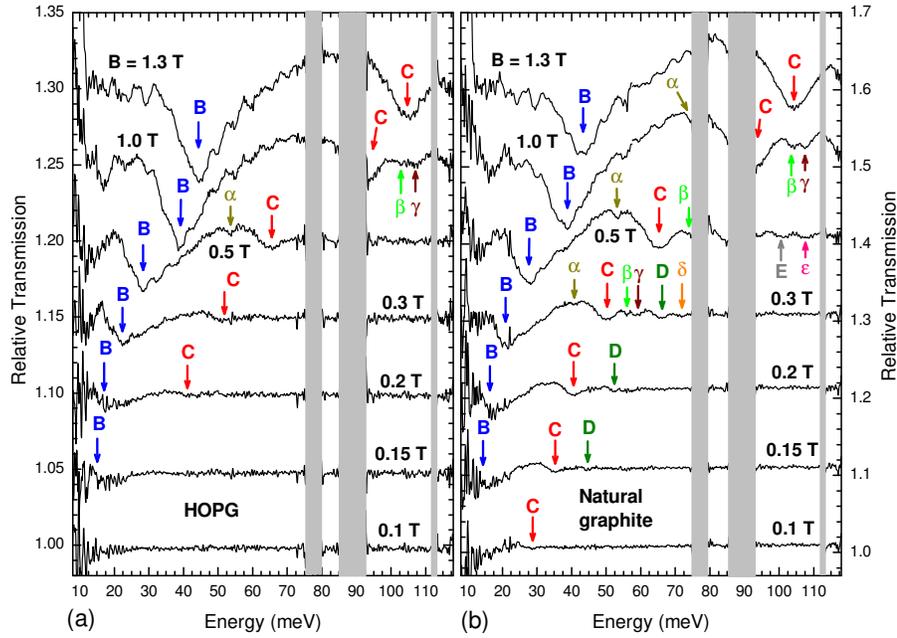}}
\end{center}
\caption{\label{LowFieldSPKT} Magneto-transmission spectra of HOPG
(a) and NG (b) taken in the spectral window 10-120 meV. The
individual spectra in (a) and (b) were shifted vertically with a
step of 0.05 and 0.1, respectively.}
\end{figure}

To explain the contradictory results obtained using different
experimental methods, among others, material properties of various
types of bulk graphite are nowadays under discussion. Namely, the
highly oriented pyrolytic graphite (HOPG) with its polycrystalline
character is assumed to exhibit more anisotropic (nearly 2D)
behaviour in comparison to Kish graphite or natural graphite (NG),
even though these materials are reported to have much better crystal
properties. Indications of such behaviour were reported e.g. in STS
experiments~\cite{MatsuiPRL05} or in transport measurements, giving
the vertical conductivity in HOPG to be much lower than in Kish
graphite~\cite{KopelevichPRL03}.

In this paper, we present  FIR magneto-transmission measurements performed on
two types of bulk graphite, HOPG and NG. We focus on spectral features
exhibiting $\sqrt{B}$-dependence, which serve as a fingerprint of Dirac-like
particles. We find no significant difference in the optical response of both
materials and evaluate the same Fermi velocity $\tilde{c}=(1.02\pm0.02)\times
10^6$~m.s$^{-1}$ as well as the pseudogap at the $H$ point $|\Delta|<10$~ meV.

\begin{figure}
\begin{center}
\scalebox{1.5}{\includegraphics*[13pt,9pt][240pt,174pt]{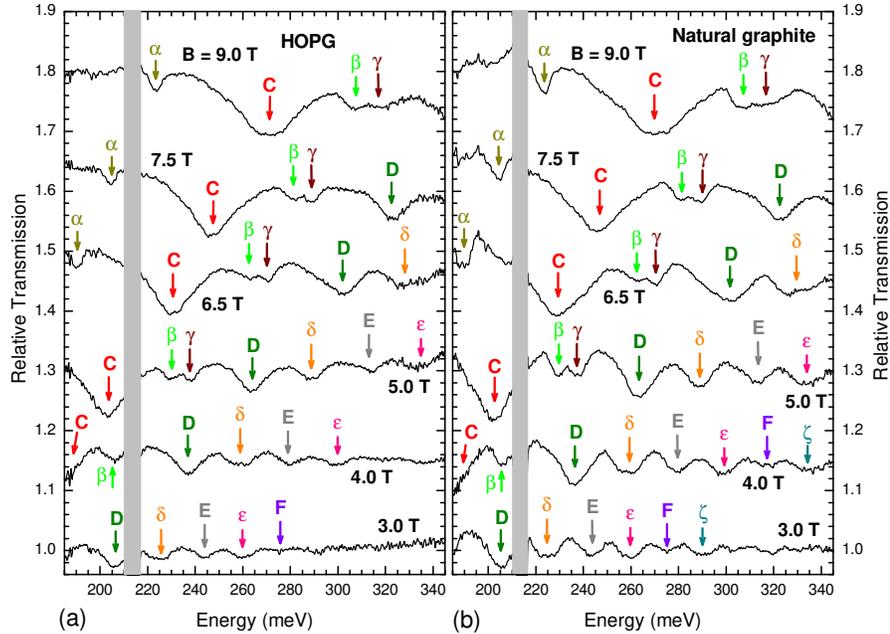}}
\end{center}
\caption{\label{MiddleFieldSPKT} Comparison of magneto-transmission spectra
taken in the spectral range 180-350 meV on HOPG and NG in the part (a) and (b),
respectively. For clarity, the successive spectra in both parts were shifted
vertically by amount of 0.15.}
\end{figure}

\section{Experiment}

Thin samples for the transmission measurements were prepared by
exfoliation of HOPG and natural graphite (NG) in the way described
in Ref.~\cite{OrlitaPRL08} and characterized using micro-Raman.
Probing the sample prepared from HOPG~\cite{OrlitaPRL08}, the
detected Raman signal corresponded to bulk
graphite~\cite{FerrariPRL06}. More complex results were obtained on
NG sample, where the shape of 2D feature in the Raman spectrum
varied scanning across both the sample, and NG crystal from which
the sample was prepared, indicating the presence of bulk graphite as
well as few-layer graphite stacks~\cite{FaugerasAPL08}. Graphite flakes covered in case
of HOPG and NG around 50\% and up to 70\% of the tape surface,
respectively. Transmission experiments were carried out on a
macroscopic circular-shaped sample with a diameter of several
millimeters.

\begin{figure}
\begin{center}
\scalebox{1.0}{\includegraphics*[15pt,15pt][309pt,245pt]{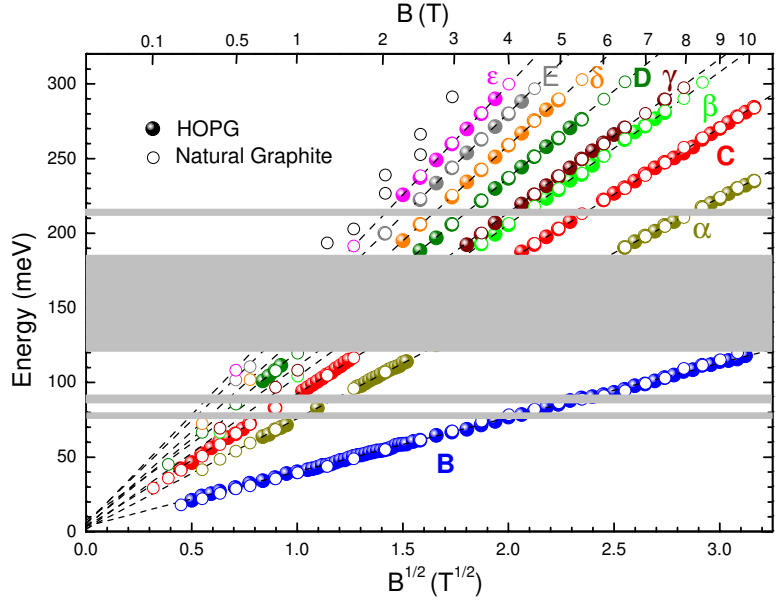}}
\end{center}
\caption{\label{FanChart} Positions of inter-LL transitions as a function of
$\sqrt{B}$ for natural graphite (open circles) and HOPG (solid circles). The
data for HOPG are taken from Ref.~\cite{OrlitaPRL08}.}
\end{figure}

The FIR experiments have been performed using the experimental setup
described in Ref.~\cite{OrlitaPRL08,SadowskiPRL06}. To measure the
transmittance of the sample, the radiation of globar, delivered via
light-pipe optics to the sample and detected by a Si bolometer,
placed directly below the sample and cooled down to a temperature of
2~K, was analyzed by a Fourier transform spectrometer. All
measurements were performed in the Faraday configuration with the
magnetic field applied along the $c$-axis of the sample. All the
spectra were taken with non-polarized light in the spectral range of
10-350~meV, limited further by several regions of low tape
transmissivity (see gray areas in Figs.~\ref{LowFieldSPKT},
\ref{MiddleFieldSPKT} and \ref{FanChart}). The transmission spectra
were normalized by the transmission of the tape and by the
zero-field transmission, thus correcting for magnetic field induced
variations in the response of the bolometer.

\section{Results and Discussion}

The FIR transmission spectra of HOPG and NG, taken at temperature of 2~K, are
presented in Figs.~\ref{LowFieldSPKT} and \ref{MiddleFieldSPKT} as a function
of the applied magnetic field. The spectra of HOPG, presented in parts (a) of
both figures, are taken from Ref.~\cite{OrlitaPRL08}. Individual absorption
lines are denoted by Roman and Greek letters, following the notation used in
Ref.~\cite{OrlitaPRL08}, and their $\sqrt{B}$-dependence is demonstrated by the
fan chart in Fig.~\ref{FanChart}. Therefore, these lines are related to the $H$
point of graphite, where the energy spectrum in the presence of the magnetic
field can be with a reasonable accuracy described by the expression:
\begin{equation}\label{LL}
E_n=\mathrm{sign}(n)\tilde{c}\sqrt{2e\hbar B |n|}=\mathrm{sign}(n)E_1\sqrt{|n|}
\qquad n=0,\pm1\pm2... ,
\end{equation}
typical of LLs in graphene. The validity of this formula is limited
to the case of the vanishing pseudogap $\Delta\rightarrow0$ at the
$H$ point, which is consistent with no splitting of the B
line~\cite{OrlitaPRL08} observed in our spectra. Assuming the LLs in
the form of~(\ref{LL}), the absorption lines B,C,D,E and F can be
assigned to the inter-LL transitions L$_{-m}\rightarrow$L$_{m+1}$
(or L$_{-(m+1)}\rightarrow$L$_m$) with $m=0,1,2,3$ and 4,
respectively. Similarly, $\alpha,\gamma,\delta,\varepsilon$ and
$\zeta$ lines can be interpreted as transitions symmetric around the
Dirac point L$_{-m}\rightarrow$L$_m$ with $m=1,2,3$ and 4,
respectively. Finally, the $\beta$ line corresponds to transitions
L$_{-1(-3)}\rightarrow$L$_{3(1)}$. Note that whereas the transitions
denoted by Roman letters have a direct counterpart in absorption
lines observed in both epitaxial and exfoliated
graphene~\cite{SadowskiPRL06,JiangPRL07}, the lines denoted by Greek
letters do not obey the graphene-like dipole selection rule
$|n|\rightarrow |n|\pm 1$, but are dipole allowed when the LL
structure around the $H$ point of bulk graphite is treated properly
within the SWM model~\cite{OrlitaPRL08,DresselhausPR65}.

Comparing Figs.~\ref{LowFieldSPKT} and \ref{MiddleFieldSPKT}, no
significant differences are found in spectra taken on both types of
bulk graphite investigated here. In both materials, the same set of
absorption lines is observed, characterized by the same positions in
the spectra, giving a Fermi velocity of
$\tilde{c}=(1.02\pm0.02)\times 10^6$~m.s$^{-1}$. The
spectra also exhibit very similar lineshapes. In NG as well as HOPG,
the B line shows no signs of splitting and thus, based on its width,
we can estimate the pseudogap to be $|\Delta|<10$~meV, in agreement
with the value of $\approx$8~meV found by Toy \emph{et
al.}~\cite{ToyPRB77}. This estimation does not support the recent
conclusion of Gr\"{u}neis \emph{et al.}~\cite{GruneisPRL08}, who
estimated the pseudogap $|\Delta|>20$~meV using the combined
approach of the ARPES experiment and calculations based on the
density-functional theory.

Data obtained on NG exhibit more pronounced spectral features, which
likely indicates a higher quality of the NG crystal in comparison to
polycrystalline HOPG. The NG sample allows us to resolve transitions
with higher LL indices at very low magnetic fields, cf. spectra at
$B=0.3$ in Fig.~\ref{LowFieldSPKT}. Similarly, the $\zeta$ line is
clearly resolved in the transmission spectra of NG and not HOPG at
higher magnetic fields.

\section{Conclusion}

To conclude, the magneto-transmission spectroscopy has been used to
probe the nature of Dirac fermions in highly oriented pyrolytic and
natural graphite. Both types of bulk graphite exhibited a very
similar, if not identical optical response, giving the Fermi
velocity $\tilde{c}=(1.02\pm0.02)\times 10^6$~m.s$^{-1}$ and the
pseudogap $|\Delta|$ below 10 meV.

\section*{Acknowledgement}

It is our pleasure to thank Yuri Latyshev for his strong interest in our work.
The present work was supported by Contract
No. ANR- 06-NANO-019, Projects No. MSM0021620834 and No.
KAN400100652, and by the European Commission through Grant No.
RITA-CT-2003-505474.

\section*{References}

\bibliography{Orlita_et_al-ref}
\end{document}